\documentclass[12pt]{iopart}

\makeatletter
\long\def\@makecaption#1#2{
  \vskip\abovecaptionskip
  \sbox\@tempboxa{#1. #2}
  \ifdim \wd\@tempboxa >\hsize
    #1. #2\par
  \else
    \hbox to\hsize{\hfil #1. #2\hfil}
  \fi
}
\makeatother

\usepackage{cite}

\usepackage{bm,amssymb}
\newcommand{\bq}{\bm{q}}
\newcommand{\bx}{\bm{x}}
\newcommand{\bn}{\bm{n}}
\newcommand{\diff}{\mathrm{d}}
\newcommand{\eqref}[1]{(\ref{#1})}
\newcommand{\order}{\mathcal{O}}

\usepackage{graphicx}% Include figure files
\usepackage{epsfig}
\usepackage{xcolor}
{} % End formatting

\begin{document}

\title[MFPT of active Brownian particles]{Mean first passage time of active Brownian particles in two dimensions}

\author{Sarafa A. Iyaniwura$^1$ and Zhiwei Peng$^2$}

\address{$^1$Vaccine and Infectious Disease Division, Fred Hutchinson Cancer Center, Seattle, Washington 98109, USA}
\address{$^2$Department of Chemical and Materials Engineering,
University of Alberta, Edmonton, Alberta T6G 1H9, Canada}

\ead{iyaniwura@aims.ac.za \& zhiwei.peng@ualberta.ca}

\vspace{10pt}
\begin{indented}
\item[]%\today 
\end{indented}

\begin{abstract}
The mean first passage time (MFPT) is a key metric for understanding transport, search, and escape processes in stochastic systems. While well characterized for passive Brownian particles, its behavior in active systems---such as active Brownian particles (ABPs)---remains less understood due to their self-propelled, nonequilibrium dynamics. In this paper, we formulate and analyze an elliptic partial differential equation (PDE) to characterize the MFPT of ABPs in two-dimensional domains, including circular, annular, and elliptical regions. For annular regions, we analyze the MFPT of ABPs under various boundary conditions.  Our results reveal rich behaviors in the MFPT of ABPs that differ fundamentally from those of passive particles. Notably, the MFPT exhibits non-monotonic dependence on the initial position and orientation of the particle, with maxima often occurring away from the domain center. We also find that increasing swimming speed can either increase or decrease the MFPT depending on the geometry and initial orientation. Asymptotic analysis of the PDE in the weak-activity regime provides insight into how activity modifies escape statistics of the particles in different geometries. Finally, our numerical solutions of the PDE are validated against Monte Carlo simulations. 
\end{abstract}

\section{Introduction}

The mean first passage time (MFPT) is a fundamental quantity in the analysis of stochastic processes and has been widely used to characterize the behavior of physical and biological systems, including diffusion, chemical reactions, and cellular processes. For passive Brownian particles, the MFPT theory is well established and extensively studied both analytically and numerically \cite{redner2001guide, gardiner2009stochastic, pillay2010asymptotic, iyaniwura2021simulation}.
Furthermore, the MFPT has been computed in a wide range of geometric domains, ranging from simple shapes such as intervals, disks, spheres, and ellipses to more complex and irregular geometries, including narrow channels, domains with small absorbing traps, and multi-scale or fractal-like structures \cite{iyaniwura2021simulation, holcman2014narrow, cheviakov2010asymptotic}.
The relative simplicity of Brownian motion allows for a rigorous treatment via an elliptic partial differential equation (PDE) that governs the MFPT under various boundary conditions.

In contrast to Brownian systems, the theory of MFPT for active  particles remains relatively underdeveloped. Active particles, which consume energy to self-propel, exhibit persistent, directionally biased motion at short times and often display effective diffusive dynamics at long times. Owing to their self-propulsion, active particles are  inherently out of thermodynamic equilibrium, often exhibiting richer dynamics than passive Brownian motion. Several theoretical models, including the active Brownian particle (ABP), run-and-tumble, and active Ornstein-Uhlenbeck processes, have been employed to study the dynamics of active particles \cite{romanczuk2012active,solon2015active,bonilla2019active,sevilla2019generalized,martin2021statistical}. The self-propulsion and model-specific properties  have made it challenging to develop a unified theoretical framework for computing the MFPT of active particles \cite{angelani2015run}.

Theoretical treatments of the MFPT for ABPs is often approached by solving the Fokker-Planck (or Smoluchowski) equation that governs the probability distribution of an ABP in its position and orientation space. With solution to the distribution, one can obtain the survival probability of a particle as a function of time, from which the MFPT can be computed as an integral. This approach, while conceptually straightforward, is computationally intensive and analytically intractable in many geometries, especially in two or three spatial dimensions \cite{maggi2014generalized, dabelow2023mfpt, iyaniwura2024asymptotic, di2023active, baouche2025first}. 

A recent study by Tanasijević et al.~\cite{tanasijevic2022microswimmers} proposed an elliptic PDE for the MFPT of ABPs, analogous to the classical MFPT equation for passive Brownian motion. Their derivation was based on the adjoint of the two-dimensional Fokker–Planck equation governing ABP dynamics~\cite{zwanzig2001nonequilibrium}. In a related development, Hillen et al.~\cite{hillen2025mean} formulated a general elliptic integro-PDE for the MFPT of velocity jump processes, and further employed parabolic scaling techniques to obtain an anisotropic effective PDE in the diffusive limit. Given the structural similarities between ABP motion and velocity jump processes, both involving persistent motion with stochastic reorientation, it is plausible that a comparable asymptotic framework could yield an elliptic PDE for ABPs.

As the theory of MFPT for active particles continues to evolve, recent advances in the field have significantly expanded our understanding of escape and target search processes for active particles in confined environments \cite{go2024active, kumar2023escape, upadhyaya2024narrow, salgado2025narrow, grebenkov2021distribution, sahoo2025target, paoluzzi2020narrow, iyaniwura2024asymptotic,baouche2024active,shee2024active,santra2020run}.
Go et al.~\cite{go2024active} analyzed the search efficiency of active particles in thermal environments, highlighting how the interplay between activity and thermal noise can optimize reaction rates with localized targets.
Similarly, Sahoo et al.~\cite{sahoo2025target} investigated target search by a polymer with an active head, demonstrating how active propulsion affects search efficiency and can lead to faster target localization depending on polymer length and activity.
Paoluzzi et al.~\cite{paoluzzi2020narrow} focused on the narrow escape time in circular domains, showing that sorting of active particles based on motility parameters is possible by tuning boundary conditions. Complementing these findings, Grebenkov et al.~\cite{grebenkov2021distribution} studied the first-reaction time distribution in shell-like geometries with partially reactive boundaries, offering a rigorous analytical framework that extends classical narrow escape theory to more realistic boundary conditions and particle dynamics. 
In a previous study~\cite{iyaniwura2024asymptotic}, we analyzed the MFPT of ABPs in one dimension under weak swimming using asymptotic methods, highlighting the influence of initial position, orientation, and swimming speed on escape times.

Several studies have specifically examined escape behaviors in constrained geometries with small openings or reactive targets. Kumar and Chakrabarti~\cite{kumar2023escape} investigated the escape dynamics of self-propelled nanorods from circular confinements with narrow exits, demonstrating the critical influence of particle shape, propulsion strength, and exit width on escape probability and time. Upadhyaya and Akella~\cite{upadhyaya2024narrow} proposed an optimal strategy for chiral active particles in narrow escape settings, identifying optimal angular velocities that minimize escape times. Salgado-García et al.~\cite{salgado2025narrow} explored the behavior of camphor-driven active particles near escape openings, highlighting facilitated escape mechanisms and aging effects, where prolonged confinement alters escape dynamics. Collectively, these studies emphasize the importance of activity, confinement geometry, particle structure, and noise in shaping the escape time distributions and optimizing target search efficiency in active systems.

In this paper, we consider an elementary random walk approach for deriving an elliptic PDE that describes the MFPT for ABPs in two- and three-dimensional domains, under specific conditions. Our method is inspired by classical techniques developed for passive Brownian motion \cite{redner2001guide, tzou2015mean, iyaniwura2021simulation}, but is adapted to capture the self-propelling dynamics inherent to active particles. Consider an ABP that self-propels at a constant speed $U_s$ along its body-fixed orientation vector $\bq$, where $\bq\cdot\bq=1$.  The orientation vector undergoes random reorientations due to rotational Brownian motion, characterized by a rational diffusivity $D_R$. In addition to rotational diffusion, the ABP undergoes translational Brownian motion with a diffusivity $D_x$. Let \( T(\bx, \bq) \) denote the expected time (or MFPT) for a particle, starting at position \( \bx \) with orientation \( \bq \), to reach the boundary of a domain $\Omega \subset \mathbb{R}^d$, where $d = 2$ or $3$. We show that $T(\bx, \bq)$ satisfies the PDE:
\begin{equation}
\label{eq:mfpt-eq}
    U_s\bq\cdot\nabla T + D_x \nabla^2 T + D_R \nabla_R^2 T=-1,
\end{equation}
where $\nabla = \frac{\partial}{\partial \bx}$ is the spatial gradient operator, and  $\nabla_R=\bq\times \frac{\partial }{\partial \bq}$ is the rotational gradient operator. A Dirichlet boundary condition (BC) is imposed at an absorbing boundary $\partial \Omega_a$: $T(\bx\in \partial \Omega_a, \bq)=0$. At a reflecting boundary $\partial \Omega_r$, we impose a Neumann BC given by $\bn \cdot\nabla T (\bx\in \partial \Omega_r, \bq)=0$, where $\bn$ is the unit normal vector at the boundary. By setting $U_s = 0$ and integrating out the orientational degrees of freedom, one recovers the classical MFPT equation for passive Brownian particles: \( D_x\nabla^2 T=-1 \), where \( T=T(\bx) \).

As expected, Eq.~\eqref{eq:mfpt-eq} is consistent with that given in Ref.~\cite{tanasijevic2022microswimmers}. By solving the PDE numerically, we study the MFPT of ABPs in several geometries in two dimensions (2D), including a disk, an annulus, and an ellipse. In the weak-activity limit, asymptotic results are developed that reveal the non-trivial effects of activity on the MFPT. The results from the PDE is validated against Monte Carlo simulations that resolve the stochastic trajectories of ABPs over time.  We further explore the influence of different boundary conditions on these domains to assess how domain structure affects the MFPT. Additionally, we examine the impact of varying swimming speed on the escape time of particles, providing insights into how self-propulsion strength shapes transport behaviors in confined environments in a non-trivial fashion. We note that, due to the high dimensionality, solving \eqref{eq:mfpt-eq} in three dimensions (3D) is challenging.

\section{Derivation of the MFPT equation via a random walk approach}

To derive Eq. \eqref{eq:mfpt-eq}, we begin with the Langevin equations that govern the stochastic trajectories of an ABP as a function of time $t$: 
\begin{equation}
\label{eq:langevin-linear}
    \frac{\diff \bx}{\diff t}= U_s \bq + \sqrt{2D_x}\;\bm{\eta},
\end{equation}
\begin{equation}
\label{eq:langevin-angular}
    \frac{\diff\bq}{\diff t} = \sqrt{2D_R}\;\bm{\xi}\times \bq.
\end{equation}
Here, $\bm{\eta}$ and $\bm{\xi}$ are independent white-noise processes. Over a small time interval \( \Delta t \), the particle translates by \( U_s\bq\Delta t \) due to self-propulsion, experiences a stochastic linear displacement \( \delta \bx \sim \mathcal{N}(0, 2D_x \Delta t  \;\mathbf{I}) \),  where $\mathbf{I}$ is the identity matrix, and changes orientation by \( \Delta \bq \sim \mathcal{N}(0, 2D_{R} \Delta t \; \mathbf{I}) \) due to rotational Brownian motion.
Defining the total displacement as $\Delta \bx = U_s \bq \Delta t + \delta \bx$, the MFPT for an ABP initially at position $\bx \in \Omega$ with orientation $\bq$ to exit the domain $\Omega$ satisfies the discrete-time random walk equation:
\begin{equation}
\label{eq:Discrte_MFPT}
T(\bx, \bq) = \Delta t +  \mathbb{E} \left[ T\left(\bx + \Delta \bx,\; \bq + \Delta \bq \right) \right]. 
\end{equation}
Similar to the backward Kolmogorov equation for a random walk in 1D \cite{tzou2015mean, iyaniwura2021simulation}, Eq.~\eqref{eq:Discrte_MFPT} describes the escape time of the particle from the domain $\Omega$ as the sum of the time  it takes the particle to make one small move, and the expected time for it to exit the domain from the new state reached after the small move. Since  \( \delta \bx \) and \( \Delta \bq \) represent stochastic (noise-driven) displacements, the state of the particle after the initial step is random. Therefore, we must average over all possible outcomes, weighted by their probability. This is captured by the expectation operator \( \mathbb{E}[\cdot] \) in the equation.

Next, we expand \( T(\bx + \Delta\bx, \bq + \Delta \bq) \) in Eq.~\eqref{eq:Discrte_MFPT} about \( (\bx, \bq) \), giving 
\begin{eqnarray}\label{Eq:T_Expand}
T(\bx + \Delta \bx, \bq + \Delta \bq) &&= T(\bx, \bq)
 + \Delta x_i \frac{\partial T}{\partial x_i}  + \Delta q_i\, \frac{\partial T}{\partial q_i}   \nonumber \\ 
&& + \frac{1}{2}  \Delta x_i \frac{\partial^2 T}{\partial x_i \partial x_j}\Delta x_j 
+  \frac{1}{2}  \Delta q_i \frac{\partial^2 T}{\partial q_i \partial q_j}\Delta q_j 
+ \cdots ,
\end{eqnarray}
where we have omitted higher order terms. We now compute the expectation of this series. Since $ T(\bx, \bq)$ is not a random variable, $ \mathbb{E}(T(\bx, \bq)) =  T(\bx, \bq)$. From the second term of the series, we have
 $\mathbb{E}[\Delta x_i] = U_s  q_i  \Delta t $ since $\mathbb{E}[\delta x_i] = 0$. The expectation of the third term vanishes due to symmetry.  
Taking the expectation of the fourth term in \eqref{Eq:T_Expand}, we have
\begin{eqnarray}
    \mathbb{E} \left[ \frac{1}{2} \, \Delta x_i \, \frac{\partial^2 T}{\partial x_i \partial x_j} \, \Delta x_j \right]
= \frac{1}{2} \, \mathbb{E} \left[ \Delta x_i \, \Delta x_j \right] 
\, \frac{\partial^2 T}{\partial x_i \partial x_j}.
\end{eqnarray}
Noting that $\mathbb{E}[\Delta x_i \Delta x_j] 
= U_s^2 \, q_i \, q_j \, (\Delta t)^2 + 2 D_x \, \Delta t \, \delta_{ij}$, we obtain
 \begin{equation}
 \label{eq:x-quadratic}
       \mathbb{E}\left[ \frac{1}{2} \Delta x_i \frac{\partial^2 T}{\partial x_i \partial x_j}\Delta x_j \right] = \left( \frac{1}{2}U_s^2\, q_i \,q_j \,\left(\Delta t\right)^2  + D_x\, \Delta t\, \delta_{ij} \right) \frac{\partial^2 T}{\partial x_i \partial x_j},
 \end{equation}
where $\delta_{ij}$ denotes the identity tensor. 
The expectation of the last term in Eq.~\eqref{Eq:T_Expand} is given by
 \begin{equation}
     \mathbb{E}\left[ \frac{1}{2}  \Delta q_i \frac{\partial^2 T}{\partial q_i \partial q_j}\Delta q_j \right]= D_R\,\Delta t\, \nabla_R^2 T,
 \end{equation}
where $\mathbb{E}[\Delta q_i \Delta q_j] = 2D_R \, \Delta t \, \delta_{ij}$ has been used. Putting everything together, we have 
\begin{eqnarray}\label{eq:Exp_Term}
\mathbb{E} \left[ T(\bx + \Delta \bx, \right. & \left. \bq + \Delta \bq) \right] =
T(\bx, \bq)
+ U_s  \bq\cdot \nabla T\,  \Delta t \nonumber \\ 
&+ D_x \nabla^2 T \, \Delta t +D_R \nabla_R^2 T\, \Delta t +\mathcal{O}[\left(\Delta t\right) ^2]
\end{eqnarray}
Inserting this into Eq.~\eqref{eq:Discrte_MFPT}, dividing by $\Delta t$, and taking the limit as $\Delta t\to 0$, we obtain the elliptic PDE for the MFPT, as given in Eq.~\eqref{eq:mfpt-eq}.

\section{MFPT in a disk}\label{sec:Disk_MFPT}

\begin{figure}
\centering 
\includegraphics[width=5in]{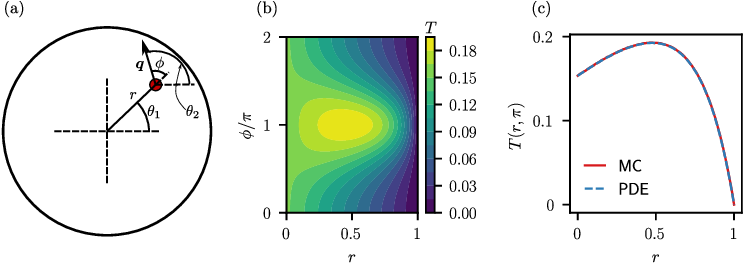}
\caption{MFPT in a disk. (a) Schematic diagram of an active Brownian particle in a disk. The radial position and angle of the particle is represented by $r$ and $\theta_1$, respectively, and its orientation angle is $\theta_2$. The angle between the radial direction and the orientation vector ($\bq$) is denoted by $\phi$. (b) Contour plot of the MFPT for the particle to escape the disk for $Pe_s=5$ and $\beta=0.1$. (c) MFPT as a function of the starting radial position $r$ for $\phi = \pi$, where the particle initially points toward the origin, computed using the PDE in \eqref{Eq:Dimless_ABPS_PDE_Polar} and Monte Carlo (MC) simulations.
}
\label{fig:fig1}
\end{figure}

In this section, we use Eq. \eqref{eq:mfpt-eq} to compute the MFPT for an ABP in a disk of radius $R$ with an absorbing boundary. Let $\bx = (r,\theta_1)$ denote the position of the ABP in polar coordinates, and let $\theta_2$ represent its orientation angle, as illustrated in Fig.~\ref{fig:fig1}a. The orientation angle is defined such that $\bq = \cos\theta_2 \, \bm{e}_x +\sin\theta_2\,\bm{e}_y$, where $\bm{e}_x$ and $\bm{e}_y$ are the unit basis vectors in $x$ and $y$ directions of a Cartesian frame, respectively. For convenience, we define the angle between the radial direction and the orientation vector $\bq$ as $\phi$; $\bq\cdot\bm{e}_r = \cos\phi$, where $\bm{e}_r$ is the unit basis vector in the radial direction. 

We write Eq.~\eqref{eq:mfpt-eq} in terms of the $(r, \theta_1, \theta_2)$ system as
\begin{eqnarray}\label{Eq:Dim_ABPS_PDE_Polar2}
U_s & \left[  \bq\cdot\bm{e}_r\frac{\partial T}{\partial r} +   \frac{\bq\cdot\bm{e}_{\theta_1}}{r} \frac{\partial T}{\partial \theta_1}   \right] \nonumber \\[1ex]
 & \qquad + D_x \left[ \frac{1}{r} \frac{\partial }{\partial r} \left( r \frac{\partial T}{\partial r} \right) + \frac{1}{r^2} \frac{\partial^2 T}{\partial \theta_1^2} \right] + D_R\,\frac{\partial^2 T}{\partial \theta_2^2} = -1,
\end{eqnarray}
where $ r \leq R$, $0 \leq \theta_1, \theta_2 \leq 2 \pi$, and $\bm{e}_{\theta_1}$ is the unit basis vector in the $\theta_1$ direction. Since we have assumed that the boundary of the disk  is absorbing, we set $T(r=R, \theta_1, \theta_2) = 0$; periodic boundary conditions are enforced in $\theta_1$ and $\theta_2$. Due to axial symmetry, it is more convenient to work with $\phi$ instead of $\theta_2$. To this end, we introduce a change of variables: $\theta=\theta_1$ and $\phi = \theta_2-\theta_1$. In terms of $\theta$ and $\phi$, we have $\partial/\partial\theta_1 = \partial/\partial\theta - \partial/\partial\phi$ and $\partial/\partial\theta_2=\partial/\partial\phi$. In the $(r, \theta, \phi)$ system, one can show that $T$ is invariant in $\theta$. We can then reduce Eq.~\eqref{Eq:Dim_ABPS_PDE_Polar2} to
\begin{eqnarray}\label{Eq:Dim_ABPS_PDE_Polar3}
U_s & \left[  \cos(\phi)\frac{\partial T}{\partial r} - \frac{\sin(\phi)}{r}  \frac{\partial T}{\partial \phi}  \right] \nonumber \\[1ex]
 & \qquad  + D_x \left[ \frac{1}{r} \frac{\partial }{\partial r} \left( r \frac{\partial T}{\partial r} \right) + \frac{1}{r^2} \frac{\partial^2 T}{\partial \phi^2}  \right] + D_R\,\frac{\partial^2 T}{\partial \phi^2} = -1,
\end{eqnarray}
with $T(r=R, \phi) = 0$ and $T(r, \phi) = T(r, \phi + 2\pi)$.

We non-dimensionalize lengths by the radius of the  disk $R$, and time with the diffusive time-scale $\tau_D :=R^2/D_x$ to obtain the dimensionless PDE given by
\begin{eqnarray}\label{Eq:Dimless_ABPS_PDE_Polar}
Pe_s &\left[  \cos(\phi)\frac{\partial T}{\partial r} - \frac{\sin(\phi)}{r}  \frac{\partial T}{\partial \phi}  \right]  + \frac{1}{r} \frac{\partial }{\partial r} \left( r \frac{\partial T}{\partial r} \right) + \frac{1}{r^2} \frac{\partial^2 T}{\partial \phi^2} + \beta\,\frac{\partial^2 T}{\partial \phi^2} = -1 \,; \nonumber \\[2ex]
 & \qquad \qquad \qquad  T(1)= 0 \,; \quad  \quad T(r, \phi) = T(r, \phi + 2\pi),
\end{eqnarray}
where  $0 < r \leq 1$,  $0 \leq  \phi \leq 2 \pi$, $Pe_s = U_s R/D_x$ is the swimming  P\'eclet number and  $\beta = \tau_D D_R$. We observe from these expression that $Pe_s$ is directly proportional to the dimensional swim speed of the particle $U_s$, and $\beta$ is directly proportional to the dimensional rotational diffusivity of the particle ($D_R$). We solve \eqref{Eq:Dimless_ABPS_PDE_Polar} numerically in \texttt{FreeFem++}~\cite{hecht2012new} using a standard finite element method (FEM). 

In Fig.~\ref{fig:fig1}b, we show a contour plot of the MFPT for the ABP to escape the disk, as a function of its initial radial position $r$ and normalized orientation $\phi/\pi$. As expected, the MFPT vanishes when the particle starts on the absorbing boundary at $r=1$. However, unlike the case of a passive Brownian particle, where the MFPT increases monotonically as the particle starts closer to the origin, the MFPT for the ABP is non-monotonic in $r$ when the initial orientation is close to $\phi = \pi$, i.e., pointing toward the origin. In this case, the MFPT increases as $r$ decreases from $r = 1$, reaches a maximum at an intermediate radius, and then decreases as $r \to 0$ (Fig.~\ref{fig:fig1}c).

To gain intuition for the non-monotonic behavior of the MFPT as the  starting position of the particle approaches the origin, we consider the time the particle spends wandering in the domain before reaching the boundary, due to its initial orientation pointing toward the origin. This wandering time initially increases as the particle starts closer to the center, since it moves inward and takes longer to eventually exit the domain. However, a maximum is reached. As the starting position moves even closer to the origin, the MFPT begins to decrease, as the particle has a higher chance of escaping the domain from the part of the boundary it is facing. In this regime, the particle spends less time wandering before reaching the boundary, resulting in a reduced MFPT. It is important to note that the MFPT decreases as the initial orientation of the particle deviates from pointing toward the origin, and the dependence on $r$ becomes more monotonic, increasing as $r$ decreases and reaching a maximum at the origin (e.g., for $\phi=0$ or $2\pi$). An ABP can exit the domain either by swimming or via translational diffusion. The interplay between these two modes of transport gives rise to non-trivial exiting dynamics shown in Fig.~\ref{fig:fig1}b, which are absent for passive Brownian particles.

A similar study was performed by Di Trapani et al.~\cite{di2023active}, offering a more detailed analysis of the escape dynamics of ABPs in a disk domain with an absorbing. Their work combines spectral analysis of the Fokker-Planck equation with Monte Carlo simulations to characterize not only the MFPT, but also the full first-passage time distribution and the associated survival probability over time. Consistent with the results shown in Fig.~\ref{fig:fig1}, their findings reveal complex anisotropic patterns in the MFPT landscape that depend on the initial position and orientation of the particle. Notably, the maximal escape times are observed away from the center of the domain, highlighting the role of the interplay between activity and geometric confinement. In contrast to their approach, the PDE we considered here directly yields the MFPT for all initial positions and orientations.

To investigate the effect of activity on the MFPT, we consider a regular perturbation expansion of the MFPT in the weak-activity regime, characterized by $Pe_s \ll 1$. In the limiting case, where  $Pe_s$ is identically zero, we recover the result for passive (i.e., Brownian) particles. For weak activity, we expand the MFPT as a regular power series, $T(r, \phi) = T_0(r,\phi) + Pe_s \, T_1(r,\phi) +\order(Pe_s^2)$. Substituting this expansion into \eqref{Eq:Dimless_ABPS_PDE_Polar}, at leading-order, we have 
\begin{eqnarray}\label{Eq:DiskLeadOrder_PDE}
&\frac{1}{r} \frac{\partial }{\partial r} \left( r \frac{\partial T_0}{\partial r} \right) + \frac{1}{r^2} \frac{\partial^2 T_0}{\partial \phi^2} + \beta\,\frac{\partial^2 T_0}{\partial \phi^2} = -1 \,; \nonumber \\[2ex]
&  T_0(1,\phi) = 0 \,; \quad  \quad T_0(r, \phi) = T_0(r, \phi + 2\pi).
\end{eqnarray}
The solution to Eq.~\eqref{Eq:DiskLeadOrder_PDE} is 
\begin{eqnarray}\label{Eq:DiskT0_Sol}
T_0(r) = \frac{1}{4}(1 - r^2),
\end{eqnarray}
which represents the MFPT for passive particles.

At $\mathcal{O}(Pe_s)$, using the solution for $T_0$, we obtain the problem for $T_1$
\begin{eqnarray}\label{Eq:DiskOrder_Pes_PDE}
 \frac{1}{r} \frac{\partial }{\partial r}& \left( r \frac{\partial T_1}{\partial r} \right) + \frac{1}{r^2} \frac{\partial^2 T_1}{\partial \phi^2} + \beta\,\frac{\partial^2 T_1}{\partial \phi^2} =  \frac{ r \cos(\phi)}{2}   \,; \nonumber \\[2ex]
 &   T_1(1,\phi)= 0 \,; \quad  \quad T_1(r, \phi) = T_1(r, \phi + 2\pi).
\end{eqnarray}
The structure of the PDE for $T_1(r,\phi)$ suggests a solution of the form $T_1(r, \phi) =A_1(r) \cos\phi $, where $A_1(r)$ satisfies the following modified Bessel equation
\begin{eqnarray}\label{Eq:Disk_A_ODE}
r^2 \frac{d^2 A_1(r)}{dr^2}  + r \frac{d A_1(r)}{dr} - (1+ r^2 \beta) A_1(r) =  \frac{ r^3}{2},
\end{eqnarray}
with the condition that $A_1(r)$ is finite as $r \to 0$ and $A_1(1)=0$, whose solution is given by
\begin{equation}\label{eq:T1-sol}
    A_1(r) =  - \frac{1}{2\beta}\left( r-\frac{I_1\left(\sqrt{\beta } r\right)}{I_1\left(\sqrt{\beta }\right)}\right), 
\end{equation}
where $I_1$ denotes the modified Bessel function of the first kind of order one.

\begin{figure}
\centering 
\includegraphics[width=5.0in]{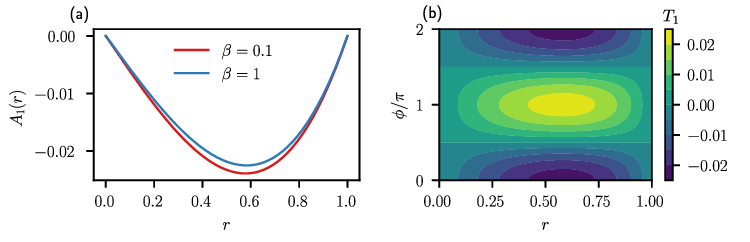}
\caption{Effect of the $\mathcal{O}(Pe_s)$ term $T_1(r, \phi)$ (Eq.~\eqref{eq:T1-sol}) on the MFPT. (a) Plots of the radial component, $A_1(r)$, as a function of the initial radial position ($r$), for different values of the dimensionless rotational diffusivity ($\beta$): $\beta=0.1$ (red curve) and $\beta=1$ (blue curve). (b) Contour plot of $T_1(r, \phi)$ with respect to $r$ and the initial orientation ($\phi$) for $\beta=1$. }
\label{fig:A1}
\end{figure}

In Fig.~\ref{fig:A1}a, we plot the radial component of the $\mathcal{O}(Pe_s)$ term in the MFPT expansion ($A_1)$, as given in Eq.~\eqref{eq:T1-sol}, as a function of the initial radial position ($r$), for different values of the dimensionless rotational diffusivity ($\beta$). For $r\in (0,1)$,  $A_1$ is negative and exhibits a non-monotonic variation as a function of $r$. When $\beta=1$, $A_1$ obtains its minimum at $r\approx 0.583$. 
From Eq.~\eqref{eq:T1-sol}, we conclude that at $\order(Pe_s)$, activity can either increase or decrease the MFPT  compared to the Brownian case (i.e., $T_0$).

 Furthermore, the increase (or decrease) in the MFPT is non-monotonic as a function of the starting radial position $r$ and orientation $\phi$. When particles initially point towards the outer absorbing boundary ($\bq\cdot\bm{e}_r=\cos\phi >0$), the MFPT decreases due to activity. The maximum reduction in the MFPT is achieved when $A_1$ attains its minimum.  Conversely, for particles initially pointing towards the disk center ($\cos\phi <0$), the MFPT increases (Fig.~\ref{fig:A1}b).
 This is because the particles must swim to the opposite side of the disk and exit there. Our analysis reveals that activity modifies the escaping dynamics in a non-trivial fashion. In particular, the initial orientation plays an important role. We further note that if the initial orientational distribution of particles is uniformly random, the MFPT at $\order(Pe_s)$ of this ensemble would vanish [see Eq.~\eqref{eq:T1-sol}]; in this case, the first correction due to activity appears at $\order(Pe_s^2)$ \cite{iyaniwura2024asymptotic}.

\section{MFPT in an annulus}

Next, we study the MFPT for an ABP in a concentric annulus. As in the disk geometry, the MFPT in the annular domain also satisfies the dimensionless PDE given in \eqref{Eq:Dimless_ABPS_PDE_Polar}. The radial interval is now $r\in[\alpha,1]$, where $\alpha = R_i/R$, with $R_i$ denoting the dimensional radius of the inner boundary. We consider an annulus with inner radius $\alpha = 0.2$, and examine different combinations of boundary conditions involving absorption and reflection at the inner and outer boundaries.

\subsection{MFPT in an annulus with absorbing inner and outer boundaries}\label{sec:Annulus_Abs_both}

We begin our analysis with the scenario in which both the inner and outer boundaries of the annular region are absorbing, i.e., $T(\alpha, \phi)=T(1,\phi)=0$. In the weak-activity regime, similar to the disk problem, we expand the MFPT as $T(r, \phi) = T_0(r,\phi) + Pe_s\, T_1(r,\phi) +\order(Pe_s^2)$. The leading-order MFPT satisfies the same PDE as the disk case given in Eq.~\eqref{Eq:DiskLeadOrder_PDE}, but with different boundary conditions. Solving this problem in an annulus, we obtain
\begin{equation}
T_0 = \frac{1-r^2}{4} + \frac{(\alpha^2-1)}{4 \ln \alpha}\ln r . 
\end{equation}
Using this solution and the PDE for $T_1$, we obtain the $\mathcal{O}(Pe_s)$ problem given by
\begin{eqnarray}\label{Eq:AnnulusOrder_Pes_PDE}
 \frac{1}{r} \frac{\partial }{\partial r}& \left( r \frac{\partial T_1}{\partial r} \right) + \frac{1}{r^2} \frac{\partial^2 T_1}{\partial \phi^2} + \beta\,\frac{\partial^2 T_1}{\partial \phi^2} 
 = \left[  \frac{r}{2} - \frac{(\alpha^2-1)}{4 \ln \alpha}\frac{1}{r}  \right] \cos\phi \,; \nonumber \\[2ex]
 &   T_1(\alpha, \phi)=T_1(1,\phi)=0 \,; \quad  \quad T_1(r, \phi) = T_1(r, \phi + 2\pi).
\end{eqnarray}
To solve \eqref{Eq:AnnulusOrder_Pes_PDE}, we assume a solution of the form $T_1(r,\phi) = A_2(r) \cos\phi$, where $A_2(r)$ satisfy the following equation
\begin{eqnarray}\label{Eq:Annulus_A2_ODE}
r^2 A_2''(r)  & + r A_2'(r) - (1+ r^2 \beta) A_2(r) =   \frac{r^3}{2} - \frac{(\alpha^2-1)r}{4 \ln \alpha}.
\end{eqnarray}
Because the analytic expression for $A_2$ is lengthy, it is provided in the supplementary material. As shown in Fig.~\ref{fig:A2}, $A_2$ is a non-monotonic function of the radial position. Unlike the disk case (cf. Fig.~\ref{fig:A1}), the sign of $A_2$ depends on the value of $r$ in an annular region. This means that for a given orientation angle, activity can either promote or hinder the escape process depending on the radial position.

\begin{figure}
\centering 
\includegraphics[width=3.0in]{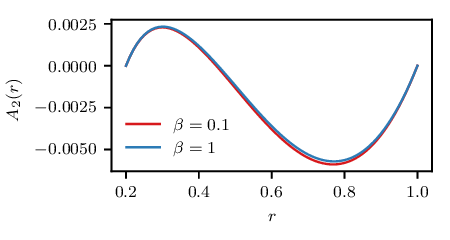}
\caption{ Plots of the radial component $A_2(r)$ of the $\order(Pe_s)$ term in the MFPT expansion for an annular region, as a function of the initial radial position ($r$), for different values of the dimensionless rotational diffusivity ($\beta$): $\beta=0.1$ (red curve) and $\beta=1$ (blue curve). Both the inner boundary (with radius $\alpha=0.2$) and outer boundary at $r=1$ are absorbing. }
\label{fig:A2}
\end{figure}

\begin{figure}
\centering 
\includegraphics[width=3.5in]{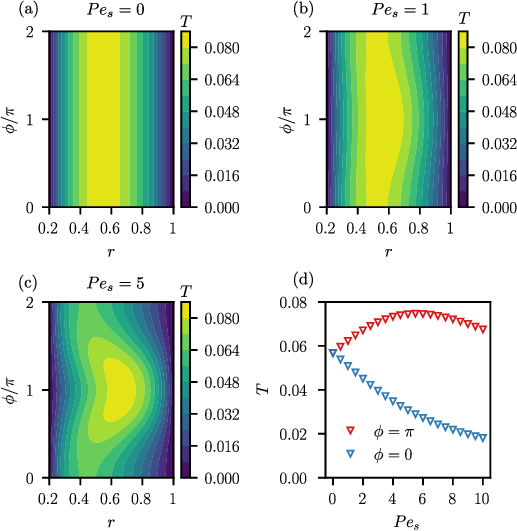}
\caption{MFPT in an annulus with absorbing inner and outer boundaries for $\alpha=0.2$ and  $\beta=0.1$. Contour plots (a--c) show numerical solutions of the PDE in Eq.~\eqref{Eq:Dimless_ABPS_PDE_Polar} for different values of $Pe_s$. (d) MFPT as a function of the Péclet number $(Pe_s)$ for $r=0.8$, $\phi=0$ (blue) and $\phi=\pi$ (red). For $\phi=\pi$, there is an optimum swim speed ($Pe_s$) that maximizes the MFPT.  }
\label{fig:annulus-bc0}
\end{figure}

To characterize the MFPT for finite $Pe_s$, we turn to numerical solutions of Eq.~\eqref{Eq:Dimless_ABPS_PDE_Polar}. In Fig.~\ref{fig:annulus-bc0}, we present contour plots of the MFPT as a function of the initial radial position ($r$) and the normalized orientation $(\phi/\pi)$ for three values of the Péclet number: $Pe_s = 0$ (no self-propulsion) (Fig.~\ref{fig:annulus-bc0}a), $Pe_s = 1$ (Fig.~\ref{fig:annulus-bc0}b), and $Pe_s = 5$ (Fig.~\ref{fig:annulus-bc0}c).
As expected, when $Pe_s = 0$, the MFPT is independent of the initial orientation, since the motion is purely diffusive. In all cases, the MFPT is lowest (i.e., $T$ vanishes) when the particle starts on either boundary, and increases as the initial position moves away from the boundaries. Owing to curvature, the maximum MFPT does not occur at the midpoint between the two boundaries, but rather closer to the inner boundary.

\begin{figure}
\centering 
\includegraphics[width=5in]{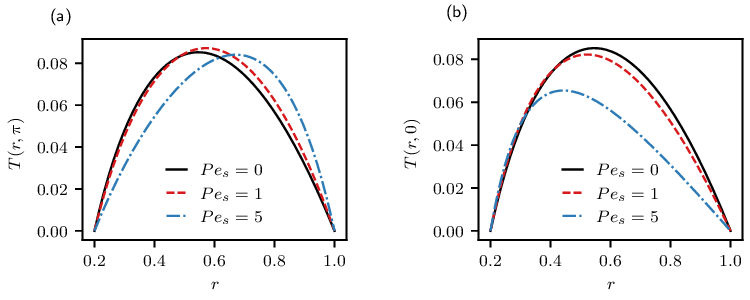}% Here is how to import EPS art
\caption{MFPT in an annulus as a function of the starting radial position for different values of $Pe_s$. Both the inner and outer boundaries are absorbing. (a) MFPT for $\phi=\pi$. (b) MFPT for $\phi=0$.  The other parameters are fixed at  $\beta=0.1$ and $\alpha=0.2$.  The line plots correspond to data at horizontal slices of the contour plots  in Fig.~\ref{fig:annulus-bc0}(a--c).}
\label{fig:annulus-bc0-T_vsr_Pes}
\end{figure}

The dependence of the MFPT on the initial orientation of the particle becomes evident as the particle starts swimming, i.e.,  $Pe_s \neq 0$, as shown in Fig.~\ref{fig:annulus-bc0}b when $Pe_s = 1$. As the swimming speed increases, the influence of orientation on the MFPT becomes more pronounced. In the contour plot for $Pe_s = 5$ (Fig.~\ref{fig:annulus-bc0}c), we observe that the location of the initial radial position where the MFPT attains its maximum shifts closer to the outer boundary, compared to the cases with $Pe_s = 0$ and $Pe_s = 1$. To better understand this shift, Fig.~\ref{fig:annulus-bc0-T_vsr_Pes} shows the MFPT as a function of radial position $r$ for the three values of $Pe_s$ considered. These results confirm that, when the particle initially points toward the inner boundary ($\phi = \pi$), the location of the maximum MFPT shifts outward as $Pe_s$ increases (Fig.~\ref{fig:annulus-bc0-T_vsr_Pes}a). Conversely, when the particle initially points toward the outer boundary ($\phi = 0$), the maximum MFPT shifts inward with increasing $Pe_s$ (Fig.~\ref{fig:annulus-bc0-T_vsr_Pes}b).

The shift in the radial position corresponding to the maximum MFPT as the swimming speed increases can be attributed to the interplay between the  initial orientation and the geometry of the domain. When the particle starts near the outer boundary and is oriented toward the inner boundary, it needs to swim towards the inner boundary before it can escape.  Despite the higher speed, the confinement effect resulting from the particle orientation and the geometry leads to an increased MFPT in this configuration. On the other hand, when the particle is initially oriented toward the outer boundary, the MFPT increases with increasing distance from that boundary, until the particle reaches a region near the inner boundary, through which it may also escape the domain. However, due to its initial orientation, the maximum MFPT decreases as its swim speed increases.

To further investigate the effect of swimming speed and initial orientation on the MFPT in this geometry, we consider the MFPT for a particle starting at radial position $r = 0.8$ with two different initial orientations: $\phi = 0$ (pointing toward the outer boundary) and $\phi = \pi$ (pointing toward the inner boundary), across a range of Péclet numbers ($Pe_s$), as shown in Fig.~\ref{fig:annulus-bc0}d. The results show that the MFPT initially increases with increasing $Pe_s$, reaches a peak, and then begins to decrease as $Pe_s$ continues to rise, when the particle is initially oriented toward the inner boundary ($\phi = \pi$, red curve). In contrast, when the particle is oriented toward the outer boundary ($\phi = 0$, blue curve), the MFPT decreases monotonically with increasing $Pe_s$.

The behavior observed for $\phi = \pi$ is consistent with our earlier interpretation of the contour plots in Fig.~\ref{fig:annulus-bc0}c: despite increased speed, the particle tends to remain in the domain longer due to the smaller surface area of the inner boundary and its initial orientation toward it. In contrast, when $\phi = 0$, the interpretation is more intuitive. Since the particle starts by pointing toward the larger outer boundary, increasing its swim speed allows it to reach and escape the boundary more quickly, thereby reducing the MFPT.

\subsection{MFPT in an annulus with a reflecting inner and an absorbing outer boundary}\label{sec:Annulus_ref_abs}

We now extend our analysis to an annular region with a reflecting inner boundary and an absorbing outer boundary.
Following the approach used to solve the annulus problem in Section~\ref{sec:Annulus_Abs_both},  we expand the MFPT in the weak-swimming regime as $T(r, \phi) = T_0(r,\phi) + Pe_s\, T_1(r,\phi) +\order(Pe_s^2)$. The leading-order MFPT is given by
\begin{equation}
T_0 = \frac{1-r^2}{4} + \frac{1}{2}\alpha^2\ln r . 
\end{equation}
At $\mathcal{O}(Pe_s)$, $T_1(r,\phi)$ satisfies
\begin{eqnarray}\label{Eq:AnnulusOrder_Pes_PDE2}
 \frac{1}{r} \frac{\partial }{\partial r}& \left( r \frac{\partial T_1}{\partial r} \right) + \frac{1}{r^2} \frac{\partial^2 T_1}{\partial \phi^2} + \beta\,\frac{\partial^2 T_1}{\partial \phi^2} 
 = \frac{1}{2} \left(  r - \frac{\alpha^2}{r}  \right)  \cos\phi \,; \nonumber \\[2ex]
 & \frac{\partial T_1}{\partial r} (\alpha, \phi)=0\,; \quad T_1(1,\phi)=0 \,; \quad T_1(r, \phi) = T_1(r, \phi + 2\pi).
\end{eqnarray}
Assuming a solution of the form $T_1(r,\phi) = A_3(r)\cos{\phi}$, we derive an equation for $A_3$ given by
\begin{eqnarray}\label{Eq:Annulus_A3_ODE}
r^2 A_3''(r)  & + r A_3'(r) - (1+ r^2 \beta) A_3(r) =   \frac{1}{2} ( r^3 - \alpha^2 r ). 
\end{eqnarray}
The analytic solution for $A_3$ is given in the supplementary material. As shown in Fig.~\ref{fig:A3}, $A_3$ is a non-monotonic function of the radial position. Similar to the disk case (see Fig.~\ref{fig:A1}), $A_2$ is negative for $r\in (\alpha, 1)$. This similarity originates from the fact that for both cases, the ABPs cannot escape from the inner boundary;  for the disk, the inner boundary is the origin. When particles point towards the absorbing boundary ($\cos\phi >0$), the MFPT is reduced for active particles.

\begin{figure}
\centering 
\includegraphics[width=3.0in]{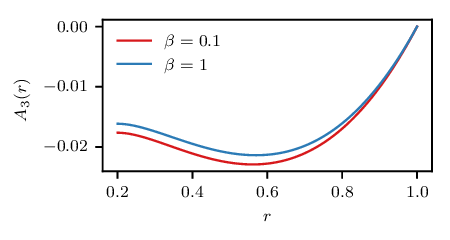}
\caption{ Plots of the radial component $A_3(r)$ of the $\order(Pe_s)$ term in the MFPT expansion for an anuular region, as a function of the initial radial position ($r$), for different values of the dimensionless rotational diffusivity ($\beta$): $\beta=0.1$ (red curve) and $\beta=1$ (blue curve). The inner boundary of the region with $\alpha=0.2$ is reflecting while the outer boundary is absorbing. }
\label{fig:A3}
\end{figure}

\begin{figure}[!h]
\centering 
\includegraphics[width=3.5in]{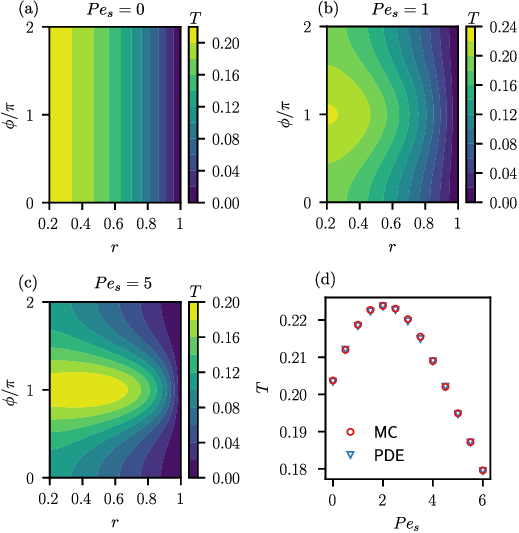}% Here is how to import EPS art
\caption{MFPT in an annular region with a reflecting inner boundary and an absorbing outer boundary for $\beta=0.1$ and $\alpha=0.2$.  Contour plots (a--c) show numerical solutions of the PDE in Eq.~\eqref{Eq:Dimless_ABPS_PDE_Polar} for different values of $Pe_s$. (d) Comparison of MFPT obtained from MC simulations and numerical solutions of the  PDE as a function of $Pe_s$ for $r=0.3$ and $\phi=\pi$. There is an optimum swim speed ($Pe_s$) that maximizes the MFPT.  }
\label{fig:annulus-bc1}
\end{figure}

As in the previous scenario with fully absorbing boundaries, we compute the MFPT numerically for ABPs to escape the domain across different Péclet numbers, as shown in Fig.~\ref{fig:annulus-bc1}. When the inner boundary is reflecting, the MFPT increases monotonically as the initial radial position of the particle moves away from the absorbing outer boundary and reaches a maximum at the reflecting inner boundary for $Pe_s = 0$ (see Fig.~\ref{fig:annulus-bc1}a). A similar trend is observed for $Pe_s = 1$, with the maximum MFPT still occurring at the inner boundary and decreasing as the initial position approaches the outer boundary.

However, for nonzero $Pe_s$, the MFPT exhibits dependence on the initial orientation of the particle. Specifically, the MFPT is highest when the particle is initially oriented toward the inner boundary ($\phi = \pi$) and decreases symmetrically as the initial orientation deviates from this direction.  For larger $Pe_s$ (see Fig.~\ref{fig:annulus-bc1}c for $Pe_s = 5$ ), the maximum MFPT occurs over a range of radial positions when the initial orientation is near $\pi$.

Comparing the contour plots in Fig.~\ref{fig:annulus-bc1}(a--c), we observe that the maximum MFPT at $\phi=\pi$ among these occurs when $Pe_s = 1$ (Fig.~\ref{fig:annulus-bc1}b), indicating that the MFPT is non-monotonic with respect to the swimming speed of the particle. In Fig. \ref{fig:annulus-bc1}d,  we plot the MFPT of an ABP starting at $r=0.3$ with an initial orientation $\phi=\pi$ as a function of $Pe_s$. For  passive particles ($Pe_s=0$), they exit the region from the outer boundary via translational diffusion. For small swim speeds, the particles tend to move towards the inner reflecting boundary, which leads to increased MFPT. In the large $Pe_s$ limit, the particles can quickly hit the inner boundary, turn back and escape from the outer boundary via swimming. This result suggests that a purely Brownian particle (i.e., $Pe_s = 0$) starting at $r = 0.3$ is more likely to escape the annular domain, bounded by a reflecting inner and absorbing outer boundary, sooner than an ABP with low swim speed and orientation toward the inner boundary. However, as $Pe_s$ increases, the ABP eventually becomes more efficient at reaching the absorbing boundary, surpassing the Brownian particle in escape likelihood. In Fig. \ref{fig:annulus-bc1}d, the results from the MFPT PDE agree with MC simulations.

\subsection{MFPT in an annulus with an absorbing inner and a reflecting outer boundary}\label{sec:Annulus_Abs_ref}

Next, we consider a similar annular region with an absorbing inner boundary at $\alpha$ and a reflecting outer boundary at $r = 1$. In this geometry, we analyze the MFPT in the weak-activity regime and expand it as $T(r, \phi) = T_0(r,\phi) + Pe_s\, T_1(r,\phi) +\order(Pe_s^2)$, where 
\begin{equation}
T_0 = \frac{1}{4}\left(\alpha^2-r^2\right) + \frac{1}{2}\left(\ln r - \ln \alpha \right).
\end{equation}
The $\mathcal{O}(Pe_s)$ MFPT satisfies
\begin{eqnarray}\label{Eq:AnnulusOrder_Pes_PDE3}
 \frac{1}{r} \frac{\partial }{\partial r}& \left( r \frac{\partial T_1}{\partial r} \right) + \frac{1}{r^2} \frac{\partial^2 T_1}{\partial \phi^2} + \beta\,\frac{\partial^2 T_1}{\partial \phi^2} 
 = \frac{1}{2} \left(  r - \frac{1}{r}  \right)  \cos\phi \,; \nonumber \\[2ex]
 &   T_1(\alpha, \phi)=0\,; \quad \frac{\partial T_1}{\partial r}(1,\phi)=0 \,;  \quad T_1(r, \phi) = T_1(r, \phi + 2\pi).
\end{eqnarray}
We assume a solution of the form
$T_1(r,\phi) = A_4(r)\cos{\phi}$, where $A_4$ satisfies the equation
\begin{eqnarray}\label{Eq:Annulus_A4_ODE}
r^2 A_4''(r)  + r A_4'(r) - (1+ r^2 \beta) A_4(r) =   \frac{1}{2} ( r^3 - r ).
\end{eqnarray}
The analytic expression for $A_4$ is given in the supplementary material. As shown in Fig.~\ref{fig:A4}, $A_4>0$ for all $\alpha \leq r \leq 1$. When particles point towards the inner (absorbing) boundary ($\cos\phi <0$), the MFPT is reduced for active particles. On the other hand, for particle pointing towards the outer (reflecting) boundary, the MFPT is elevated compared to the passive case.

\begin{figure}
\centering 
\includegraphics[width=3.0in]{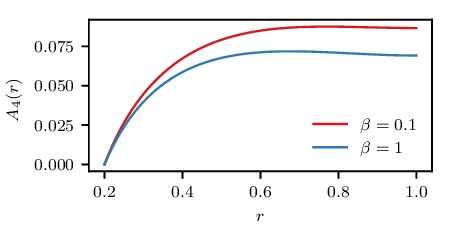}
\caption{Plots of the radial component $A_4(r)$ of the $\order(Pe_s)$ term in the MFPT expansion for an annular region, as a function of the initial radial position ($r$), for different values of the dimensionless rotational diffusivity ($\beta$): $\beta=0.1$ (red curve) and $\beta=1$ (blue curve). The inner boundary of the region with $\alpha=0.2$ is absorbing while the outer boundary  is reflecting. }
\label{fig:A4}
\end{figure}

\begin{figure}
\centering 
\includegraphics[width=3.5in]{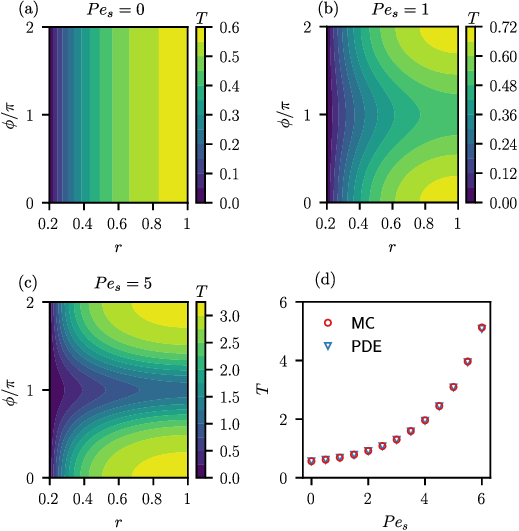}% Here is how to import EPS art
\caption{ MFPT in an annular region with an absorbing inner boundary and a reflecting outer boundary  for $\beta=0.1$ and $\alpha=0.2$. Contour plots (a--c) show numerical solutions of the PDE in Eq.~\eqref{Eq:Dimless_ABPS_PDE_Polar} for different values of $Pe_s$.  (d) Comparison of MFPT obtained from MC simulations and numerical solutions of the PDE as a function of $Pe_s$ for $r=0.9$ and $\phi=0$. As the swim speed increases, the MFPT becomes longer.  }
\label{fig:annulus-bc2}
\end{figure}

As in the previous analyses, we compute the MFPT for different Péclet numbers. Contour plots of the MFPT for this configuration are shown in Fig.~\ref{fig:annulus-bc2}(a–c). Since the inner boundary is absorbing, the MFPT is lowest near the inner boundary and highest near the reflecting outer boundary. Interestingly, unlike in the previous scenarios, the maximum MFPT occurs when the initial orientation of the particle points towards the outer boundary, and it decreases symmetrically as the orientation deviates from that direction for $Pe_s \ne 0$. Furthermore, the minimum MFPT for nonzero $Pe_s$ is observed when the particle initially points towards the absorbing inner boundary ($\phi = \pi$).

From the contour plots in Figs.~\ref{fig:annulus-bc2}(a–c), we observe that the maximum MFPT in the annular region increases as the swimming speed of the particle increases. This is another counterintuitive result, as one might expect the time required for an ABP to exit the domain to decrease with increasing swimming speed. 

We hypothesize that this increase in MFPT is due to the small circumference of the absorbing inner boundary, which provides a narrower escape region compared to the reflecting outer boundary. To further investigate this effect, we consider an ABP initially located at $r_0 = 0.9$ and oriented towards the outer boundary ($\phi = 0$). We compute the MFPT for swimming speeds ranging from $Pe_s = 0$ to $Pe_s = 6$. The results confirm that the MFPT increases with increasing swimming speed in this configuration.

\section{MFPT in an ellipse}

\begin{figure}[!h]
\centering 
\includegraphics[width=5in]{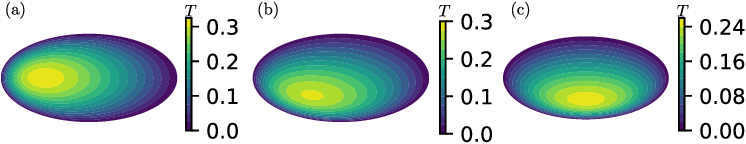}% Here is how to import EPS art
\caption{MFPT in an ellipse with $\beta=b^2D_R/D_x=0.1$, $Pe=U_sb/D_x=5$, and $a/b=2$, where $a$ and $b$ are the semi-major and semi-minor axes, respectively. Contour plots (a--c) show the MFPT for different values of the orientation angle $\theta_2$, where $\bq\cdot\bm{e}_x=\cos\theta_2$. (a) $\theta_2=0$. (b) $\theta_2=\pi/4$. (c) $\theta_2=\pi/2$. The location of the maximum MFPT is off-center.}
\label{fig:ellipse-contour}
\end{figure}

Lastly, we study the MFPT for an ABP in an elliptical region. In our analysis, we consider an elliptical region with axis ratio $a/b = 2$, where $a$ and $b$ denote the semi-major and semi-minor axes, respectively. The MFPT is computed by solving Eq.~\eqref{eq:mfpt-eq} numerically using FEM, as illustrated in Fig.~\ref{fig:ellipse-contour}. Unlike the disk and annular geometries [cf. Eq.~\eqref{Eq:Dimless_ABPS_PDE_Polar}], Eq.~\eqref{eq:mfpt-eq} in an elliptical region involves three mathematical dimensions.

Due to the symmetry of the elliptical region about both its major and minor axes, we consider three representative initial orientations:
\begin{itemize}
    \item $\theta_2 = 0$: pointing along the semi-major axis toward the boundary,
    \item $\theta_2 = \pi/4$: pointing diagonally between the semi-major and semi-minor axes,
    \item $\theta_2 = \pi/2$: pointing along the semi-minor axis toward the boundary.
\end{itemize}

The results in Fig.~\ref{fig:ellipse-contour} show the MFPT for an ABP to escape the elliptical region, starting from all positions in the domain. Compared to the cases where the initial orientation is either $\theta_2 = 0$ (Fig.~\ref{fig:ellipse-contour}a) or $\theta_2 = \pi/4$ (Fig.~\ref{fig:ellipse-contour}b), the MFPT for $\theta_2=\pi/2$
has the smallest maximum. This is  due to the shorter length of the semi-minor axis, as a particle initially pointing toward the boundary in this direction is more likely to reach and exit the domain sooner than those oriented along longer trajectories.

Interestingly, for all three orientations, the maximum MFPT does not occur at the center of the region but rather at a location opposite the initial orientation, symmetric about the center of the domain. For instance, when $\theta_2 = 0$, the maximum MFPT is attained at a point along the major axis on the opposite side of the domain. These results highlight the interplay between the swimming persistence of ABPs and the geometry in dictating the escape dynamics.

\section{Discussion}

In this paper, we employ an elliptic PDE alongside Monte Carlo simulations to analyze the MFPT for an ABP escaping from bounded 2-D domains with various geometries and boundary conditions. Our analysis reveals a range of interesting and sometimes counterintuitive behaviors, particularly in how the MFPT depends on the initial position, orientation, and swimming speed, quantified by the Péclet number ($Pe_s$).

For a disk domain, the MFPT exhibited non-monotonic dependence on the initial radial position when the ABP was oriented towards the origin. Specifically, starting closer to the origin initially increases the MFPT due to longer time spent wandering near the center, but beyond a certain point, the MFPT begins to decrease as the particle can more readily escape in the opposite direction. As the initial orientation deviated from pointing directly inward, the MFPT became more monotonic, increasing with decreasing radial distance and reaching a maximum at the origin, similar to the behavior observed in passive Brownian particles \cite{iyaniwura2021simulation, tzou2015mean}.

In annular domains, we considered several boundary conditions. When both the inner and outer boundaries were absorbing, we found that the MFPT was lowest at the boundaries and highest at interior points, with the maximum occurring off-center. The location of the maximum shifted depending on the swim speed and initial orientation. For low swim speeds, the maximum occurred closer to the inner boundary due to its smaller surface area, but for higher $Pe_s$, the maximum continues to shift towards the outer boundary as the orientation of the particle come closer to pointing directly inward.
This behavior reflects the interplay between boundary geometry and directional persistence in the motion of ABPs.

In the case of a reflecting inner boundary and an absorbing outer boundary, the MFPT increased monotonically as the initial radial position moved inward, with a clear maximum at the inner boundary when $Pe_s=0$. Interestingly, at intermediate swim speeds, the MFPT reached a higher maximum than at both lower and higher swim speeds. This non-monotonic dependence on swimming speed is counterintuitive, as one might expect faster particles to always escape more quickly. However, due to the presence of a reflecting boundary, swimming particles can spend more time near the wall, which results in a delay in its escape.

When the boundary conditions were reversed, the MFPT was lowest at the inner boundary and highest at the outer boundary, as expected. However, we again observed an increase in MFPT with increasing $Pe_s$ for particles starting near the outer boundary and pointing inward. This behavior is driven by the small surface area of the absorbing inner boundary and the particle’s persistent inward orientation, causing it to spend more time exploring the domain before finding the exit. This underscores how geometry and orientation interact to determine escape statistics for ABPs.

Finally, in elliptical domains, we observed that the MFPT was strongly dependent on the initial orientation of the particle relative to the axes. The maximum MFPT was largest when the particle was oriented along the semi-major axis and smallest when oriented along the semi-minor axis. Furthermore, for each orientation, the maximum MFPT occurred at a point symmetric to the orientation direction about the center. These observations highlight the symmetry inherent in ABP dynamics within regular but anisotropic domains and demonstrate the importance of domain geometry in shaping escape behavior.

While the elliptic PDE and accompanying simulations provide valuable insights into the MFPT of an ABP in confined domains, several simplifications in the current framework present opportunities for further exploration. The analysis assumes idealized boundary conditions and neglects hydrodynamic interactions, which may play a significant role in more realistic microscale settings. Moreover, the model considers an ABP with constant speed and rotational diffusion, omitting effects such as motility fluctuations and external fields. The present study focuses primarily on two-dimensional domains, with no attention paid to the richer geometric and dynamical complexities inherent in three-dimensional spaces. 

Nevertheless, these limitations suggest several promising research directions. Future work will extend the framework to complex domains, where the coupling between geometry and particle orientation is expected to reveal new MFPT phenomena. Additionally, asymptotic analysis of the elliptic PDE in various limits may yield analytical approximations and a deeper understanding of the system behavior. Another direction involves exploring multiple-trap problems in confined domains, where escape is only possible through small, localized traps (see \cite{iyaniwura2021optimization, iyaniwura2021simulation, iyaniwura2021asymptotic},  and therein for related problems on passive particles). Such configurations give rise to complex search dynamics and pose challenging optimization questions.
A well-known biological example of the localized trap problem is a sperm cell searching for the egg in the female reproductive tract, where the egg acts as a small absorbing target in a large domain~\cite{friedrich2007chemotaxis}.
Developing perturbation theory for these problems, based on the derived elliptic PDE, could provide analytical insight into escape times and trap interactions in the small-trap limit.

Overall, our results demonstrate that ABPs exhibit rich and varied escape dynamics that are highly sensitive to domain geometry, boundary conditions, swimming speed, and initial orientation. These findings have implications for understanding the transport and confinement of microswimmers in structured environments and may inform the design of microfluidic devices for targeted delivery and sorting of active particles.

\ack 
Z. P. was supported by the Natural Sciences and Engineering Research Council of Canada (NSERC) through the Discovery Grants program, under Grant No. RGPIN-2025-05310.

\section*{References}  % Manually insert heading
\bibliographystyle{iopart-num}  % or iopart-num-long or iopart-num-short
\bibliography{references}             % refs.bib is your .bib file

\end{document}